\documentstyle[aas2pp4]{article}

\def\simless{\mathbin{\lower 3pt\hbox 
     {$\rlap{\raise 5pt\hbox{$\char'074$}}\mathchar"7218$}}}   %< or of order 
\def\simmore{\mathbin{\lower 3pt\hbox 
     {$\rlap{\raise 5pt\hbox{$\char'076$}}\mathchar"7218$}}}   %> or of order 
\received{} 
\accepted{} 
\slugcomment{Submitted to Ap.J, 8 Jan 2000} 
 
\lefthead{Reig et al.} 
\righthead{Lags in GRS 1915+105} 
 
\begin{document} 
 
\title{Phase lag variability associated with the 0.5--10 Hz QPO \\
in GRS~1915+105} 
 
\author{ 
P.~Reig\altaffilmark{1,2},
T. Belloni\altaffilmark{3},
M.~van~der~Klis\altaffilmark{4}, 
M. M\'endez\altaffilmark{4,5},
N. D. Kylafis\altaffilmark{1,2},
E. C. Ford\altaffilmark{4}
}
 
\altaffiltext{1}{Foundation for Research and Technology-Hellas, 711 10  
Heraklion, Crete, Greece} 
 
\altaffiltext{2}{Physics Department, University of Crete, 710 03 Heraklion, 
Crete, Greece} 

\altaffiltext{3}{Osservatorio Astronomico di Brera, Via E. Bianchi 46, I-23807 
Merate (LC), Italy}
 
\altaffiltext{4}{Astronomical Institute ``Anton Pannekoek'', 
       University of Amsterdam and Center for High-Energy Astrophysics, 
       Kruislaan 403, NL-1098 SJ Amsterdam, the Netherlands} 
       
\altaffiltext{5}{Facultad de Ciencias Astron\'omicas y Geof\'{\i}sicas, 
Universidad Nacional de La Plata, paseo del Bosque S/N, 1900 La Plata, 
Argentina}
 
\begin{abstract} 
 
We have used {\em Rossi  X-ray  Timing  Explorer}  data to measure the
lags between  soft (2--5  keV) and hard  (5--13  keV)  photons  and to
study the aperiodic variability of the superluminal black hole candidate
GRS 1915+105 during low-flux states.  The power density spectra exhibit 
quasi-periodic oscillations (QPO) whose frequency increases with
increasing count rate and varies in the  frequency  range 0.6--8 Hz.  A 
correlation  between the QPO frequency  and the phase lag spectra is
reported  for the first time.  This correlation is found for both the
phase lag continuum and the phase lag at the QPO  frequency.  We find that
as the QPO  frequency  moves to higher  values the phase  lags  reverse 
sign  from  positive  to  negative.  The absolute value of the lag always
increases with photon energy.  The negative  (soft) lags are  associated 
with a softer energy spectrum,  whereas the positive (hard) lags are seen
when the source is harder.  We describe a possible scenario that may
account for the change in the sign of the lags.

\end{abstract} 
 
 
\section{Introduction} 
 
The  X-ray  source  GRS~1915+105  was  discovered  in 1992  with the 
WATCH instrument   onboard  GRANAT   (Castro-Tirado  et  al.  1992). 
Since  {\em RXTE} established  continuous  coverage in 1996 February, 
GRS~1915+105 has never switched  off  completely.  It  was  the  first 
galactic  object  to  show superluminal  radio  expansion  (Mirabel  \& 
Rodr\'{\i}guez  1994).  It is located at a distance between 6 to 12.5 kpc
(Mirabel \& Rodr\'{\i}guez  1994; Fender  et al.  1999).  Due to high 
extinction  in the  line of  sight  no optical  study  of GRS  1915+105 
has been  possible  and a near infrared counterpart  has been  suggested 
(Mirabel  et al.  1994; Bo\"er et al. 1996).  No binary mass function or
orbital period are known.

The  source  is  believed  to  host  a  black  hole  on  the  basis  of its
similarities  to the  dynamically  proven  black hole  system GRO  1655--40
(Orosz  \&  Bailyn  1997)  and its  X-ray  luminosity,  which is above  the
Eddington limit for a neutron star.  The X-ray spectrum of  GRS~1915+105 is
also  typical  of black  hole  candidates,  requiring  two main  components
(Belloni et al.  1997a):  a power-law  component which accounts for the high
energy part of the  spectrum  and a thermal  component  (multi-color  black
body) which  dominates at energies below 10 keV (Mitsuda et al.  1984).  At
times, only the power law  component is observed  (Trudolyubov,  Churazov, \&
Gilfanov 1999; Belloni et al.  2000).

GRS~1915+105  displays  a  remarkable  richness  in  variability  (see
e.g. Greiner, Morgan \& Remillard 1997).  Belloni et al.  (1997a,b) 
modeled the main type of  variations  observed  in this source as due to
the onset of a thermal-viscous  instability.  In this framework, the
intervals of low flux correspond  to the  non-observability  of  the 
innermost  portion  of  the accretion  disk.   During three long (30-100 
days) periods in 1996 and 1997, no strong flux  variability  was observed
from  GRS~1915+105.  On the basis of an X-ray color  analysis of the {\em
RXTE}/PCA  data, Belloni et al. (2000)  interpreted  these quiet periods
as the result of the recovery from an  instability  that  affected  a
larger  portion of the  accretion  disk, combined  with a relatively 
lower value of the  accretion  rate.  Detailed timing  analysis  of part
of  these  observations  has  been  presented  by Trudolyubov et al.
(1999).

Three different types of QPOs have been identified in GRS~1915+105  (Morgan,
Remillard \& Greiner 1997).  At higher frequencies a sharp ($Q \approx$ 20)
QPO with  constant  centroid  frequency  at 67 Hz and whose  rms  amplitude
increases with photon energy has been detected.  This QPO has been proposed
to arise in the  inner  accretion  disk and its  frequency  is  thought  to
reflect  the  properties  of the  black  hole.  At much  lower  frequencies
(0.001--0.1  Hz),  the  source  occasionally  shows  high-amplitude  QPOs or
brightness  "sputters".  These  variations  probably  correspond  to a disk
instability.  The third class of QPOs  observed  in  GRS~1915+105  comprises
those  detected in the frequency  range 0.5--10 Hz.  They are linked to the
properties  of the  accretion  disk  since  their  centroid  frequency  and
fractional  rms have been  reported  to  correlate  with the  thermal  flux
(Markwardt,  Swank, \& Taam 1999;  Trudolyubov  et al.  1999) and  apparent
temperature  (Muno, Morgan, \& Remillard  1999) of the disk.  These QPOs are
observed only during the low-flux intervals when, according to the model of
the  thermal-viscous  instability, a portion of the inner accretion disk is
not observable.

A phase-lag  analysis of the observation  which showed the strongest signal
at 67 Hz was  presented  by Cui (1999):  the 67 Hz QPO shows a marked  hard
lag,  while  the low  frequency  (well  below  1 Hz) QPO  shows  a  complex
hard/soft lag structure.

In this  paper,  we  concentrate  on the  phase  lags  associated  with  the
properties  of the  0.5--10  Hz QPO during two of the long quiet  intervals
mentioned above.

\section{Observations} 
 
The data used in this  work  were  retrieved  from the  public  {\em 
RXTE} archive and  correspond to two different  sets of  observations. 
The first set took place  between  1996 July 11 and 1996 August 18 and
consists of 27 observation intervals (OBI).  The second set  comprises 16 
observation intervals covering the period between  1997   September  29 
and  1997  October  24.  These  two  sets  of observations  correspond to
low-flux  intervals in which the source did not display  structured 
variability  either  in the  PCA  or in the  ASM  data (Belloni et al. 
2000).  The journal of the {\em RXTE}  observations  is given in
Table~\ref{observ}.  Due to its low-Earth orbit, {\em RXTE} data consist
of a  number  of  contiguous  data  intervals   (typically  1  hour 
long) interspersed with  observational gaps produced by Earth occultations
of the source and passages of the satellite  through the South  Atlantic 
Anomaly. Each observation interval   was  analysed   separately  and 
appear  as  different entries  in  Table~\ref{observ}.  A suffix 
$a, b, c ,...$  has been added to the data set identifiers accordingly.

\section{Analysis and Results} 

\subsection{Power density spectrum}
 
In order to study the  source  variability,  we
divided the 2--60 keV PCA light  curves (no energy  selection  was made) of
each  observation  into 16 s segments  and  calculated  the  Fourier  power
spectrum of each  segment.  The Nyquist  frequency was 256 Hz for the first
set of  observations  and 64 Hz for the  second.  The  contribution  by the
photon counting noise was computed and subtracted from each power spectrum.
The power  spectra  were  normalized  such that  their  integral  gives the
squared  fractional rms variability  (Belloni \& Hasinger 1990).  The power
spectra were then averaged  together to produce one power spectrum for each
observation interval.  

In all power spectra a strong  band-limited  noise  component  was present,
together with one or more QPO peaks.  We fitted the power  spectra  using a
model  that  consisted  of a broken  power law  and two  Lorentzian  peaks,  
representing  the band-limited noise
continuum  and the QPO plus its second  harmonic,  respectively.  While all
power  spectra  clearly  showed  one QPO peak, in some  cases the  second
harmonic was not required.  Generally, the power spectra are  characterized
by a flat power-law below the QPO frequency ($\alpha$=0.1--0.3) and a steep
power-law   above  the  QPO   ($\alpha$=   2.7--3.0).  The   slope  of  the
low-frequency  power-law  tends to increase with  increasing QPO frequency.
While the QPO  frequency  increases  with count rate, its  amplitude  (rms)
decreases  (Fig~\ref{cs_qpo_rms}).  Trudolyubov  et al.  (1999)  found  the
same trend  during  the 1996  November--1997  April  low-flux  period.  The
results of the power spectral fitting are given in Table~\ref{observ}.

\subsection{Phase lags}

For each  observation,  we produced a phase-lag  spectrum.  We divided 
the data into two energy bands (soft:  2--5 keV; hard:  5--13 keV) and
produced a cross  spectrum  for  every  16 s.  The  cross  spectrum  is 
defined  as $C(j)=X_1^*(j) X_2(j)$,  where  $X_1$  and  $X_2$  are the 
complex  Fourier coefficients  for  the  two  energy  bands  at  a 
frequency  $\nu_j$,  and $X_1^*(j)$  is the  complex  conjugate  of 
$X_1(j)$  (van  der Klis et al. 1987).  The phase lag  between  the 
signals  in the two  bands at  Fourier frequency  $\nu_j$ is 
$\phi_j=arg[C(j)]$  (the position angle of $C(j)$ in the complex plane)
and the corresponding time lag $\phi_j/2\pi  \nu_j$.  We calculated an
average cross vector $C$ by averaging the complex values over multiple
adjacent 16 s spectra,  and then finding the final value of $\phi$ vs. 
frequency.  The error in $\phi$ is  computed  from the observed variance
of $C$ in the real and imaginary  directions.  We limited our  phase lag
analysis  to  frequencies  below 10 Hz, well  below the  region  where
binning  effects  are  important  (Crary  et al.  1998).  In all  phase
lag spectra,  positive  lag values  mean that hard  photons  are  lagging 
soft photons.  No  correction  for  dead-time  effects in the phase lag 
spectra (cross-talk  between energy channels) was done since this effect
was found to be negligible.

In order to study the  variability  of the phase lags and its 
relationship with the  properties  of the QPO we obtained the mean phase
lag  spectrum  by averaging the cross-correlation vectors in four
different QPO frequency ranges, $\nu_{QPO} < 1$  Hz,  $1  \leq  \nu_{QPO} 
< 2$  Hz, $2  \leq  \nu_{QPO}  < 3$ Hz  and $\nu_{QPO}  \geq 3$ Hz. 
Figure~\ref{pds_lag}  shows  these  four phase lag spectra  together with
a  representative  power spectrum of the group.  One can see  that  both
the lag at the QPO  frequency,  and the  characteristic shape of the lag
continuum vary  drastically.  A common  feature of the lag spectra is the
presence of a broad dip at around the QPO frequency.  As the QPO
frequency  increases,  its position relative to the dip changes.  The
frequency of the dip also varies although not as fast: at low QPO
frequencies (below 2 Hz) the QPO occurs slightly before the dip minimum,
at QPO frequencies above 2 Hz it occurs at around the dip minimum. At 
QPO frequencies $\simmore 3.5$ Hz a second dip at lower Fourier
frequency develops in the phase lag spectra.

At low values of the QPO  frequency, the continuum of the phase lag
spectrum is curved and positive  whereas at high and intermediate QPO
frequencies the continuum is flatter and close to zero lag.  Averaging
over the 0.06--0.6 Hz range, that is, below any of the QPOs,  for  the 
four  cases  shown  in  Figure~\ref{pds_lag}  we find  the following 
values of the phase lags:  0.215$\pm$0.004  rad for  $\nu_{QPO}$ $<$  1 
Hz,   0.159$\pm$0.004   rad  for  1  $\leq   \nu_{QPO}   <$  2  Hz,
0.071$\pm$0.004 rad for 2 $\leq \nu_{QPO} <$ 3 Hz and --0.024$\pm$0.003
rad for $\nu_{QPO} \geq$ 3 Hz.

We also derived phase lags of the  continuum and at the QPO  frequency 
for each  observation,  by measuring the average vector in the frequency 
range 0.1--5 Hz and $\nu_{QPO} \pm  \frac{1}{2} \Delta \nu$,
respectively.  $\Delta \nu$ is the FWHM of the QPO peak.  The results 
were  plotted as a function of  QPO  frequency   and  are  shown  in  
Figure~\ref{freq_lag}.  A  tight correlation  between  these  quantities 
is  found.  As the  QPO  frequency (or count rate) increases the lag
decreases, changing sign at around 2 Hz. The opposite behavior was
reported for GX 399--4 by Nowak et al. (1999). These authors found the
shortest time delays at the lowest flux observation.  The increase above 5
Hz observed in Figure~\ref{freq_lag} (bottom) is due  to the dip moving
toward higher frequencies.  This  change  in the sign of the lag means 
that at a certain point soft photons lag hard photons.  Similar results
were obtained for the lags measured at the theoretical harmonics of the
QPO frequency.  The frequency at which the lags change from positive to
negative are consistent with 4, 6 and 8 Hz, that is, 2, 3 and 4 times that
of the QPO (2 Hz).

In  Figure~\ref{lag_ener},  the  dependence  of the phase  lags (at the QPO
frequency)  on energy is plotted.  Different  symbols  correspond  to three
different values of the QPO frequency:  squares for  $\nu_{QPO}$=0.999  Hz,
dots for  $\nu_{QPO}$=2.316  Hz and  stars  for  $\nu_{QPO}$=4.68  Hz.  The
energy bands shown are 2.0--5.0, 5.4--6.9, 7.2--9.4 and 9.8--13.1 keV.  The
phase lag was obtained by taking the softest band as the reference band and
by computing a  cross-power  spectrum  between each of the energy bands and
the reference  band.  At low  frequencies,  the phase lags are positive and
increase with energy; at high  frequencies  the phase lags are negative and
decrease  with energy; at  intermediate  frequencies  the lags are zero and
roughly constant.

We also calculated  background subtracted light curves corresponding to
the following  bands in  pulse-height  channels:  0--13  (2.0-5.0 keV,
band A), 14--35  (5.0-13.0  keV,  band B) and  36-100  (13.0-60.0  keV, 
band C) and produced hard and soft colors as HR1 = B/A and HR2 = C/A,
respectively.  In Figure~\ref{hr1_lag}  we have plotted the  phase lags of
the continuum (0.1--5 Hz) and the QPO frequency as a function of HR1.  The
negative lags occur when the source is softer (HR1  $\simless$ 1.2), 
whereas  the hard lags are seen when the source  spectrum  is harder (HR1
$\simmore$ 1.3).  A similar trend is obtained if the hard color HR2 is
used.  The anti-correlation between the QPO frequency and HR1 is a
consequence of the anti-correlation of the colors with count rate. 

It is worth noting that no soft component is required to fit the energy
spectrum in the observations presented here with the possible exception of
the ones with the highest QPO frequency. During low-flux states in GRS
1915+105 the X-ray spectra are dominated by the power-law component.
Therefore, variations in HR1 correspond to variation in the power-law
index. For a more detailed   interpretation  of X-ray colors in terms of
spectral models, see Belloni et al.  (2000).

\section{Discussion} 
 
We have analysed {\em RXTE} data of the superluminal source GRS~1915+105 during
two quiescent periods and measured lags of the 5--13 keV photons  relative
to the 2--5 keV photons. We have found that both positive and negative
lags are present. Belloni et al.  (1997a,b) have shown that the complex
spectral and temporal variability  of GRS~1915+105 may be  explained  by
the rapid  removal  and subsequent  refilling  of the  inner  region  of
the  accretion  disk.  The quiescent  state  corresponds to the absence of
this inner region and it is characterized  by a relatively low count rate
($\simless$ 20000 {\em RXTE} PCA c/s), power-law dominated X-ray spectrum and
the presence of 0.5--10 Hz QPO.   

In this state both the timing (QPO frequency, lags) and spectral
parameters (hardness ratios, temperature of disk) correlate (or
anticorrelate) with count rate. Thus a selection according to QPO
frequency implicitly implies a selection on spectral hardness or count
rate. Of special interest here are the properties of the 0.5--10 Hz QPO
(centroid, amplitude), which are strongly  correlated  with the changes of
spectral and timing  parameters and can be summarized as follows:  as the
frequency  of the QPO  increases  {\em i)} the  flux of the soft 
component (disk) increases (Markwardt et al.  1999), {\em ii)} the
temperature of the accretion disk increases (Muno et al.  1999), {\em
iii)} the fractional rms amplitude of the QPO decreases (Trudolyubov et
al.  1999).  In this work we find that {\em iv)} as the  frequency of the
QPO  increases the lag between hard and soft  photons  decreases, 
changing  sign for  $\nu_{QPO} \simmore 2$ Hz (Fig. ~\ref{freq_lag}), {\em
v)} negative lags occur when the power-law  spectrum  is soft and 
positive  lags  when  it is hard  (Fig. \ref{hr1_lag}),  {\em  vi)} the 
0.5--10  Hz QPO  disappears  when  the PCA intensity is high ($\simmore$
20000 counts per second).

The X-ray  spectra of black hole systems  consist of a power-law 
component, thought to be caused by Comptonization, plus a soft thermal 
component, which is modeled as a multi-temperature blackbody coming from
an optically thick accretion disk.  The relative  strength of these
components varies with  X-ray  flux in the 1--10 keV band. In the standard
Comptonization model only hard, i.e. positive lags, are expected. The
initially low-frequency photons scatter off energetic electrons and gain
energy. The larger the number of scatterings the higher the energy of the
escaping photons  and also the longer the time they spend in the
Comptonizing region.

In what follows we discuss  a simple scenario which might, at least {\em
qualitatively}, account for the soft, i.e negative lags.  Let us consider
a  Comptonizing  region in which the  Comptonization  process becomes 
more  efficient  closer to the black  hole. One  such situation  could 
be, for example, a disk consisting of cold electrons with a Keplerian 
distribution of velocities (bulk Comptonization) or a hot corona with a
temperature gradient, the inner parts being hotter than the outer parts
(thermal Comptonization). Let us examine the Keplerian case. Near the
black hole, the rotational  velocity of the electrons is close to the 
speed of light and those  photons  that find themselves in this
region get up-scattered due to bulk-motion  Comptonization. Note than in
this model the Comptonizing region is the accretion disk itself. The
source of soft photons can be inside (the disk itself) or outside the
disk. What is needed is that some photons find their way to the
high-velocity parts of the disk. Preliminary results of this type of
Comptonization have been reported by Kylafis \& Reig (1999).

If the optical depth is small, the majority of the input photons  escape
after a few scatterings. In this case the photons that undergo some
scatterings (hard photons) lag soft photons.  But if the optical depth is
large, the soft input  photons undergo a random-walk  through the medium
prior to escape and gain energy. Some of these now hard photons find 
themselves  in the outer parts of the Comptonizing  region where the
velocity of the electrons is low and they begin to give away energy by
means of the direct Compton effect. We expect then that soft photons lag
hard photons. In order for such  downward  scattering  to occur the
optical depth must be large and the disk must be geometrically thick so
that the photons  sample a large fraction of the  Comptonizing  medium.

If we accept that the optical depth gives a measure of the mass accretion
rate and that, in turn, larger accretion rates imply higher count rates,
then this simple model may also explain the dependence of the lag sign on
intensity. Positive lags occur when the count rate is low, whereas
negative lags are detected for high count rates (this observational fact
can be inferred from Fig~\ref{cs_qpo_rms} and \ref{freq_lag}). Low
intensity, that is low mass accretion rate, implies low optical depth, hence
up-scattering only or equivalently, positive lags. High intensity implies
large accretion rates or large optical depths and therefore diffusion plus
down-scattering, opening up the posibility of negative lags.

Given the relatively short time lags (of the order of a few milliseconds)
the Comptonizing region does not need to be very large (a few tens of
gravitational radii, $r_g=GM/c^2$). Note also, that the negative time lags
are typically an order of magnitude shorter than the positive lags.

\acknowledgements

This work was  supported  by the  Netherlands  Foundation  for  research
in astronomy  (ASTRON) under grant  781-76-017,  by the  Netherlands 
Research School for Astronomy  (NOVA), and the NWO Spinoza grant 08-0 to
E.P.J.  van den Heuvel.  One of us (PR)  acknowledges  support from the
European  Union through the Training  and  Mobility  of Researchers 
Network  Grant  ERBFMRX/CT98/0195.  This research has made use of data
obtained through the High Energy Astrophysics Science   Archive  
Research  Center  Online   Service,   provided  by  the NASA/Goddard Space
Flight Center.

\clearpage 
%---------------------------------------------------------------------------- 
\begin{deluxetable}{lccccccc} 
\scriptsize 
\tablecolumns{10} 
\tablecaption{RXTE observations of GRS 1915+105 \label{observ}} 
\tablewidth{0pt} 
\tablehead{} 
\startdata 
OBI	& MJD	&PCA$^a$    &$\nu_{QPO}$  & rms$^b$ (\%)  &rms$^b$ (\%) &lag (rad)  
& lag (ms)  \\
	&	&(c/s)	&(Hz) 	      &(0.01-20 Hz)&(QPO) &at $\nu_{QPO}$ 
& at $\nu_{QPO}$ \\
\hline 
&&&&&&&\\
\multicolumn{8}{c}{1996 July 11 -- August 18} \\ 
&&&&&&&\\
\hline 
 10408-01-22-00a     &50275.0905   &9778.3 &  3.468$\pm$0.011     &9.58$\pm$0.15 
   &9.17$\pm$0.25   &	 -0.082$\pm$0.015  & -3.8$\pm$0.7  \\
 10408-01-22-00b     &50275.1072   &9789.0 &  3.475$\pm$0.005     
&11.09$\pm$0.07   &9.31 $\pm$0.15  &	 -0.096$\pm$0.009  & -4.4$\pm$0.4\\
 10408-01-22-01      &50275.2238   &9282.4 &  2.768$\pm$0.006     
&10.45$\pm$0.10   &10.68 $\pm$0.13 &	 -0.045$\pm$0.006  & -2.6$\pm$0.3\\
 10408-01-22-02a     &50275.3572   &9008.1 &  2.561$\pm$0.005     
&10.45$\pm$0.09   &11.39$\pm$0.12  &	 -0.027$\pm$0.006  & -1.7$\pm$0.4\\
 10408-01-22-02b     &50275.4238   &9175.3 &  2.840$\pm$0.023     &9.31$\pm$0.10 
   &9.96$\pm$0.87   &	 -0.032$\pm$0.034  & -1.8$\pm$1.9\\
 10408-01-23-00a     &50278.4921   &9735.2 &  3.480$\pm$0.004     &9.87$\pm$0.08 
   &10.27$\pm$0.12  &	 -0.098$\pm$0.007  & -4.5$\pm$0.3\\
 10408-01-23-00b     &50278.5585   &9709.6 &  3.614$\pm$0.005     
&10.20$\pm$0.07   &9.90$\pm$0.11   &	 -0.103$\pm$0.006  & -4.6$\pm$0.3\\
 10408-01-23-00c     &50278.6266   &10457.7 &  4.209$\pm$0.009    &9.65$\pm$0.08 
   &9.12$\pm$0.11   &	 -0.122$\pm$0.006  & -4.6$\pm$0.2\\
 10408-01-24-00a     &50280.1702   & 8923.9 &  2.238$\pm$0.005    
&11.45$\pm$0.10   &12.13$\pm$0.11  &	 -0.001$\pm$0.006  &  -0.1$\pm$0.4\\
 10408-01-24-00b     &50280.2266   &8900.8 &  2.316$\pm$0.005     
&10.95$\pm$0.12   &12.00$\pm$0.16  &	 0.001$\pm$0.005  &  0.1$\pm$0.4\\
 10408-01-24-00c     &50280.2933   &8955.6 &  2.537$\pm$0.004     
&11.62$\pm$0.15   &11.34$\pm$0.15  &	 -0.036$\pm$0.006  & -2.3$\pm$0.4\\
 10408-01-24-00d     &50280.3599   &9011.1 &  2.598$\pm$0.008     
&10.68$\pm$0.15   &10.95$\pm$0.28  &	 -0.055$\pm$0.011  & -3.3$\pm$0.7\\
 10408-01-25-00a     &50283.4947   &8263.6 &  1.110$\pm$0.003     
&12.92$\pm$0.11   &11.67$\pm$0.20  &	 0.172$\pm$0.007  & 24.7$\pm$1.0\\
 10408-01-25-00b     &50283.5620   &8154.0 &  1.077$\pm$0.003     
&13.06$\pm$0.10   &11.24$\pm$0.21  &	 0.185$\pm$0.007  & 27.3$\pm$1.0\\
 10408-01-25-00c     &50283.6286   &8076.9 &  1.184$\pm$0.002     
&12.28$\pm$0.14   &12.06$\pm$0.16  &	 0.154$\pm$0.006  & 20.6$\pm$0.8\\
 10408-01-27-00a     &50290.5774   &7887.2 &  0.644$\pm$0.002     
&13.73$\pm$0.11   &9.74$\pm$0.28   &	 0.247$\pm$0.014  & 72.7$\pm$4.1\\
 10408-01-27-00b     &50290.6322   &7925.8 &  0.621$\pm$0.003     
&13.12$\pm$0.12   &9.66$\pm$0.22   &	 0.285$\pm$0.010  & 73.0$\pm$2.7\\
 10408-01-27-00c     &50290.6988   &7826.5 &  0.629$\pm$0.003     
&13.04$\pm$0.20   &10.28$\pm$0.19  &	 0.272$\pm$0.008  & 68.9$\pm$2.1\\
 10408-01-28-00a     &50298.5696   &7751.1 &  0.999$\pm$0.003     
&12.33$\pm$0.11   &11.36$\pm$0.18  &	 0.189$\pm$0.006  & 30.0$\pm$1.0\\
 10408-01-28-00b     &50298.6362   &7755.6 &  0.962$\pm$0.003     
&13.49$\pm$0.12   &11.21$\pm$0.21  &	 0.201$\pm$0.007  & 33.3$\pm$1.2\\
 10408-01-28-00c     &50298.7029   &7533.5 &  0.928$\pm$0.003     
&13.16$\pm$0.13   &11.55$\pm$0.19  &	 0.221$\pm$0.007  & 38.0$\pm$1.2\\
 10408-01-29-00a     &50305.3736   &7951.5 &  1.673$\pm$0.004     
&12.51$\pm$0.13   &11.98$\pm$0.17  &	 0.085$\pm$0.007  &  8.1$\pm$0.6\\
 10408-01-29-00b     &50305.4398   &8070.8 &  1.858$\pm$0.004     
&12.15$\pm$0.11   &11.97$\pm$0.16  &	 0.049$\pm$0.005  &  4.2$\pm$0.5\\
 10408-01-29-00c     &50305.5064   &8078.7 &  1.962$\pm$0.003     
&12.16$\pm$0.11   &11.95$\pm$0.15  &	 0.036$\pm$0.005  &  3.0$\pm$0.4\\
 10408-01-30-00a     &50313.3135  &11560.3  &  4.551$\pm$0.009    &8.75$\pm$0.10 
   &8.95$\pm$0.08   &	 -0.140$\pm$0.006  & -4.9$\pm$0.2\\
 10408-01-30-00b     &50313.3801  &12979.6  &  5.030$\pm$0.008    
&10.85$\pm$0.08   &7.26$\pm$0.22   &	 -0.171$\pm$0.006  & -5.4$\pm$0.2\\
 10408-01-30-00c     &50313.4468  &14701.7  &  5.551$\pm$0.015    
&13.21$\pm$0.13   &4.72$\pm$0.18   &	 -0.188$\pm$0.008  & -5.4$\pm$0.2\\
\hline
&&&&&&&\\
\multicolumn{8}{c}{1997 September 29 -- October 24} \\
&&&&&&&\\ 	
\hline  	
 20402-01-48-00a     &50720.5910  &21914.4   &  7.56 $\pm$0.03   &18.09$\pm$0.15 
   &2.50 $\pm$0.06  &	 -0.260$\pm$0.011  & -5.5$\pm$0.2\\
 20402-01-48-00b     &50720.6572  &12618.6   &  4.68 $\pm$0.01   &11.83$\pm$0.06 
   &5.62 $\pm$0.09  &	 -0.147$\pm$0.008  & -5.0$\pm$0.3\\
 20402-01-48-00c     &50720.6783  &11805.7   & 4.54$\pm$0.04     &11.45$\pm$0.20 
   &7.14$\pm$0.29   &	 -0.179$\pm$0.022  & -6.3$\pm$0.8\\
 20402-01-49-00a     &50729.3275  & 8714.6   &  2.911$\pm$0.005  &10.18$\pm$0.07 
   &10.89$\pm$0.11  &	 -0.033$\pm$0.006  & -1.8$\pm$0.3\\
 20402-01-49-00b     &50729.3940  & 8369.3   &  2.621$\pm$0.006  &10.60$\pm$0.10 
   &11.05$\pm$0.16  &	 -0.015$\pm$0.007  &  -0.9$\pm$0.4\\
 20402-01-49-01      &50730.3949  & 8197.8   &  2.721$\pm$0.005  &10.76$\pm$0.07 
   &10.96$\pm$0.12  &	 -0.041$\pm$0.006  & -2.4$\pm$0.3\\
 20402-01-50-00      &50735.5474  & 6579.2   &  0.844$\pm$0.003  &13.51$\pm$0.18 
   &10.52$\pm$0.29  &	 0.239$\pm$0.010  & 45.0$\pm$2.0\\
 20402-01-50-01a     &50737.4047  & 6607.5   &  1.016$\pm$0.003  &13.76$\pm$0.11 
   &10.82$\pm$0.21  &	 0.201$\pm$0.009  & 31.5$\pm$1.3\\
 20402-01-50-01b     &50737.4766  & 6373.4   &  1.079$\pm$0.003  &13.60$\pm$0.14 
   &11.09$\pm$0.22  &	 0.189$\pm$0.010  & 27.8$\pm$1.4\\
 20402-01-51-00a     &50743.2947  & 6626.3   &  1.399$\pm$0.006  &13.92$\pm$0.17 
   &11.71$\pm$0.31  &	 0.130$\pm$0.012  & 14.7$\pm$1.3\\
 20402-01-51-00b     &50743.3377  & 6602.3   &  1.355$\pm$0.004  &13.73$\pm$0.12 
   &11.60$\pm$0.20  &	 0.135$\pm$0.007  & 15.9$\pm$0.8\\
 20402-01-51-00c     &50743.4092  & 6584.0   &  1.399$\pm$0.005  &13.84$\pm$0.13 
   &11.38$\pm$0.20  &	 0.133$\pm$0.007  & 15.1$\pm$0.8\\
 20402-01-51-00d     &50743.4801  & 6640.4   &  1.442$\pm$0.003  &13.55$\pm$0.14 
   &11.97$\pm$0.20  &	 0.116$\pm$0.008  & 12.8$\pm$0.9\\
 20402-01-52-00a     &50746.5509  & 6657.3   &  1.413$\pm$0.005  &13.77$\pm$0.12 
   &11.27$\pm$0.22  &	 0.134$\pm$0.009  & 15.2$\pm$1.1\\
 20402-01-52-00b     &50746.6190  & 6667.5   &  1.482$\pm$0.005  &13.86$\pm$0.14 
   &11.89$\pm$0.24  &	 0.149$\pm$0.010  & 16.1$\pm$1.1\\
 20402-01-52-00c     &50746.6898  & 6767.9   &  1.602$\pm$0.004  &12.90$\pm$0.18 
   &11.86$\pm$0.27  &	 0.083$\pm$0.010  &  8.3$\pm$1.0\\
\tableline 
\enddata
\tablenotetext{a}{In the range 2-60 keV and 5 PCU.}
\tablenotetext{b}{Normalization according to Belloni \& Hasinger (1990).}
\end{deluxetable} 
%---------------------------------------------------------------------------- 
 
\clearpage 

\figcaption[fig1.ps]{
Variation of the $rms$ amplitude and frequency of the QPO with count rate.
\label{cs_qpo_rms}
}

\figcaption[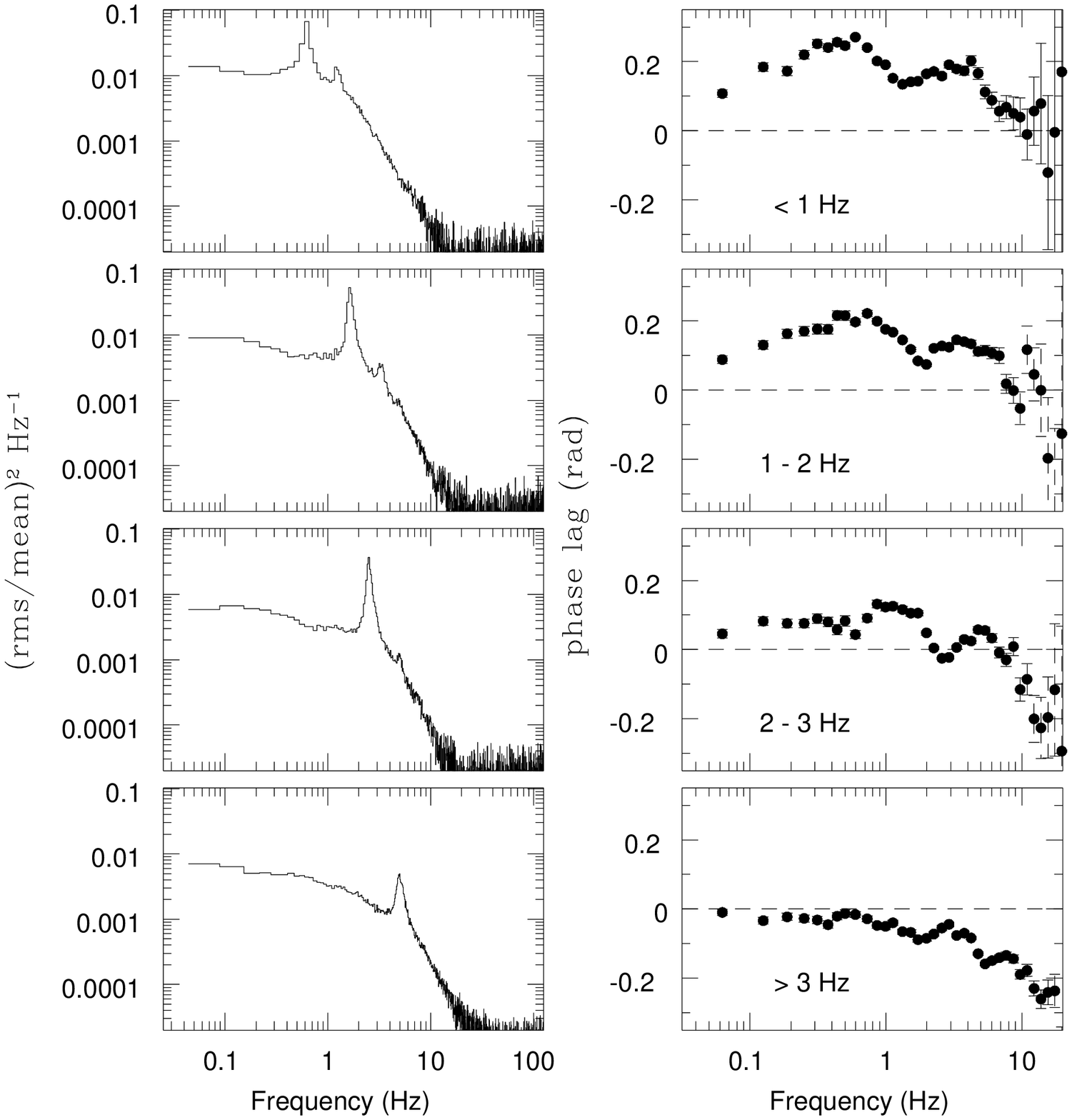]{
Power and phase lag spectra versus Fourier frequency for four QPO frequency
ranges.  The phase lag  spectra  are the average of all those  included  in the
frequency  range  shown.  The power  spectra  are  individual  examples of a
representative member of the group. Positive lags mean that hard photons lag
soft photons.
\label{pds_lag} 
} 
 
\figcaption[fig3.ps]{ 
Phase lag at the QPO frequency (top panel) and phase lag of the continuum
in the frequency range 0.1--5  Hz  (bottom panel) versus QPO  frequency. 
Different  symbols represent  different  observing runs as
follows:  1996 July 11  -- August 18 (circles), 1997 September 29
--October 24 (stars) 
\label{freq_lag} 
} 

\figcaption[fig4.ps]{
Dependence of the lag (at the QPO frequency) on energy.  Different symbols
correspond to three different values of the QPO frequency: squares for 
$\nu_{QPO}$=0.999 Hz, dots for $\nu_{QPO}$=2.316
Hz and stars for $\nu_{QPO}$=4.68 Hz. The lags were calculated for the
energy bands 5.4--6.9 keV, 7.2--9.4 kev and  9.8--13.1 keV with respect to
the soft band  2.0--5.0 keV.
\label{lag_ener}
}

\figcaption[fig5.ps]{ 
Phase lag and QPO frequency as functions of the soft color HR1=5--13
keV/2--5 keV. Negative lags are seen only when the source is soft.
\label{hr1_lag}
}

\end{document}